\documentclass[11pt, a4paper]{article}

\usepackage{anysize}
\marginsize{3cm}{3cm}{1.5cm}{1.5cm}

\usepackage{amsfonts}
\usepackage{amssymb}
\usepackage{graphics}
\usepackage{epsfig}
\usepackage{graphicx}
\usepackage{epsfig}
\usepackage{amssymb}
\usepackage{amsfonts}

\usepackage{amsmath}
\usepackage{xcolor}
\usepackage{float}
\usepackage{enumerate}
\usepackage{slashed}
\usepackage{hyperref}
\hypersetup{colorlinks=true,linkcolor=blue,citecolor=blue} 

\numberwithin{equation}{section}

\newcommand {\beq} {\begin{equation}}
\newcommand {\eeq} {\end{equation}}
\newcommand{\bea}{\begin{eqnarray}}
\newcommand{\eea}{\end{eqnarray}}
\newcommand{\bit}{\begin{itemize}}
\newcommand{\eit}{\end{itemize}}
\def\nl{\nonumber \\}
\def\Tr{{\rm Tr}}

\def\a{\alpha}

\def\b{\beta}

\def\s{\sigma}

\def\p{\partial}

\def\le{\left(}
\def\ri{\right)}

\def\beq{\begin{equation}}
\def\eeq{\end{equation}}

\def\th {\tilde{h}}

\begin{document}

\begin{titlepage}

\begin{flushright}

\end{flushright}
\bigskip
\begin{center}
{\LARGE  {\bf
Nonrelativistic trace and diffeomorphism  \\ 
anomalies in  particle number background
  \\[2mm] } }
\end{center}
\bigskip
\begin{center}
{\large \bf  Roberto  Auzzi$^{a,b}$},
 {\large \bf Stefano Baiguera$^{c}$} {\large \bf and } {\large \bf Giuseppe Nardelli$^{a,d}$} 
\vskip 0.20cm
\end{center}
\vskip 0.20cm 
\begin{center}
$^a${ \it \small  Dipartimento di Matematica e Fisica,  Universit\`a Cattolica
del Sacro Cuore, \\
Via Musei 41, 25121 Brescia, Italy}
\\ \vskip 0.20cm 
$^b${ \it \small{INFN Sezione di Perugia,  Via A. Pascoli, 06123 Perugia, Italy}}
\\ \vskip 0.20cm 
$^c${ \it \small{Universit\`a degli studi di Milano Bicocca and INFN, 
Sezione di Milano - Bicocca, \\ Piazza
della Scienza 3, 20161, Milano, Italy}}
\\ \vskip 0.20cm 
$^d${ \it \small{TIFPA - INFN, c/o Dipartimento di Fisica, Universit\`a di Trento, \\ 38123 Povo (TN), Italy} }
\end{center}
\vspace{3mm}

\begin{abstract}
Using the heat kernel method,
we compute nonrelativistic trace anomalies for Schr\"odinger
theories in flat spacetime, with a generic background gauge field
for the particle  number symmetry, both for a free scalar and a free fermion.
The result
is genuinely nonrelativistic, and it has no counterpart in the relativistic case.
 Contrary to the naive expectations, the anomaly is not gauge-invariant; this is similar 
 to the non-gauge covariance of the non-abelian relativistic anomaly. 
We  also show that, in the same background, 
the gravitational anomaly for a nonrelativistic scalar vanishes.
\end{abstract}

\end{titlepage}

\section{Introduction}

Trace anomaly gives  
powerful constraints on the possible degrees of freedom
which can emerge in the infrared of a strongly coupled 
relativistic and unitary theory: 
Zamolodchikov $c$-theorem in $d=2$ \cite{Zamolodchikov:1986gt}
and the $a$-theorem in $d=4$ \cite{Cardy:1988cwa,Osborn:1989td,Jack:1990eb,Osborn:1991gm,Komargodski:2011vj,Komargodski:2011xv}
give us examples of monotonically decreasing quantities between 
the UV and the IR conformal fixed points. 
For condensed matter applications, it would be interesting to 
generalize such results to the nonrelativistic case.  

Newton-Cartan (NC) geometry was originally introduced as a covariant formulation of
Newtonian gravity. In the last few years, it found several 
applications in condensed matter systems such as 
quantum Hall effect and fermions at unitarity, see {\it e.g.} 
 \cite{Son:2005rv,Hoyos:2011ez,Son:2013rqa,Geracie:2014nka}.
 For theories with Schr\"odinger invariance,
NC gravity  provides a natural set of sources for the operators
in the energy-momentum tensor multiplet.

Promising candidates for nonrelativistic $a$-theorems
are given by type-A trace anomalies \cite{Deser:1993yx}.
In the case of  Schr\"odinger-invariant theories 
coupled to NC geometry
in $2+1$ dimensions, a natural candidate for a monotonically decreasing
 $a$-function was  introduced in \cite{Jensen:2014hqa}
and further studied in \cite{Auzzi:2015fgg,  Arav:2016xjc, Auzzi:2016lrq,
Auzzi:2016lxb,Auzzi:2017jry,Fernandes:2017nvx, Pal:2017ntk}. 
 All these works 
 (with the exception of \cite{Fernandes:2017nvx})
 assume that the trace anomaly is 
 invariant under diffeomorphisms, Milne and $U(1)$ gauge transformations.
The case of Lifshitz theories was studied {\it e.g.} in
\cite{Adam:2009gq, Baggio:2011ha, Griffin:2011xs, Arav:2014goa,
Barvinsky:2017mal, Arav:2016akx,Pal:2016rpz} and no natural candidate
for a monotonically decreasing type-A term
 in the trace anomaly was found so far.

In general, anomalies correspond to violations of current conservation in the presence
of background fields.
In the simplest incarnations, such as the ABJ chiral anomaly \cite{Adler:1969gk,Bell:1969ts}, 
they can be written as gauge-invariant functions. 
Subsequently, Bardeen \cite{Bardeen:1969md} showed that anomalies may not
be gauge-covariant: gauge invariance for background fields
might be formally lost in the regularization procedure.
Such gauge violating terms are not just a feature of chiral anomalies; they can appear 
also in trace anomalies \cite{Keren-Zur:2014sva}. 
As discussed in \cite{Auzzi:2015yia}, 
the presence of these terms in supersymmetric theories 
is instrumental in deriving $a$-maximization
\cite{Intriligator:2003jj}
using Osborn's local renormalization group formalism 
\cite{Osborn:1989td,Jack:1990eb,Osborn:1991gm}.

The first explicit calculation of the Schr\"odinger trace anomaly in NC background 
was performed in \cite{Auzzi:2016lxb} for the scalar case and in \cite{Auzzi:2017jry} for the fermionic case.
There, the heat kernel procedure was used and 
 a NC background with vanishing particle number gauge potential was chosen.
With this assumption,  the trace 
anomaly result was  gauge and Milne boost invariant.
Subsequently, Ref.~\cite{Fernandes:2017nvx} did a related calculation in the scalar case 
using the Fujikawa method with a NC background with a non-vanishing $U(1)$ 
particle number\footnote{More correctly, this $U(1)$ symmetry in the presence of different 
species of fields $\psi_i$ corresponds to the mass, because in the minimal coupling it enters the action as
$-\sum_i m_i A_0 |\psi_i |^2$, where $m_i$ is the mass of the field $\psi_i$. In the presence of a single species, mass
and particle number are proportional to each other. For simplicity, we refer to this $U(1)$ symmetry as particle number.}
 gauge field.
Surprisingly, the trace anomaly was not $U(1)$ gauge invariant.

In this paper we investigate such violations with the heat kernel formalism.
We consider the cases of a free non relativistic scalar 
and  a free fermion in $2+1$
dimensions and we will compute
 the expectation value of the
trace of the energy-momentum tensor for a flat background
geometry and a generic source $A_\mu$ for the particle number.

Moreover, Ref.~\cite{Fernandes:2017nvx}  also found 
a non-vanishing diffeomorphism anomaly, both in the presence of space-time curvature and $U(1)$
gauge field. Given the recent interesting applications of diffeomorphism anomaly in condensed matter systems, see {\it e.g.} \cite{Stone:2012ud,Gromov:2014gta,Can:2014awa}, one of the motivation of this work is also to deepen and understand the nature of the this anomaly in nonrelativistic theories.
We compute this anomaly in the presence of a background gauge field and, surprisingly,  we find a vanishing result. As we shall explain,  this is not necessarily in contradiction with 
\cite{Fernandes:2017nvx}: we find that 
it is possible define a ``subtracted''  energy-momentum tensor which is conserved, in the spirit of \cite{Bardeen:1984pm}.
 Therefore, the apparent difference may be all due to a different renormalization and subtraction procedure.

This paper is organized as follows: in section
\ref{pre} we introduce the notation and the sources for 
the background currents. In section \ref{tratra} we compute the trace anomaly
in a particle number background both for a nonrelativistic 
boson and a fermion. In section \ref{diffe} we discuss diffeomorphism anomaly
for a nonrelativistic scalar and we compare the results with \cite{Fernandes:2017nvx}.  
We conclude in section \ref{conclu} whereas
technical details are deferred to  appendices.


\section{Preliminaries and notation}
\label{pre}

We will consider a non relativistic free scalar and  fermion in $2+1$
 dimensions coupled to a NC background geometry. 
As a useful method to deal in a convenient way with all
the space-time symmetries, we will use the null-reduction
trick \cite{Duval:1984cj} from an extra-dimensional relativistic 
$3+1$ dimensional theory.
Useful reference about NC geometry formalism  include
\cite{Jensen:2014aia,Hartong:2014pma,Hartong:2014oma,
Hartong:2015wxa,Geracie:2016bkg,Banerjee:2014nja,Banerjee:2015rca,Banerjee:2016laq}.

Since we are dealing with fermions, we will need both
curved space-time indices and tangent space ones for the frame field.
The index conventions that we shall use are as follows:
late latin capital  indices ({\it e.g.} $ M, N, \ldots $),  denote  $3+1$ dimensional curved space-time indices;
 early latin capital indices  ({\it e.g.} $A, B, \ldots , $) denote tangent space indices, whose metric is locally flat. 
The coordinate  $x^-$ corresponds to the null-reduction direction.
The extra-dimensional indices of the curved space and of
 the locally flat tangent space are as follows:
 \beq
\begin{aligned}
&  M=(-,\mu)=(-,+,i)    \qquad (i=1,2)  \\
&  A=(-,\alpha)=(-,+,a)  \qquad  (a=1,2)  \, ,
\end{aligned}
\eeq
where $+$ denotes the nonrelativistic time direction and $i$, $a$
the space ones (for the curved, tangent space case, respectively).

The NC space-time geometry is described by
  a positive definite symmetric rank $2$ tensor 
$h^{\mu \nu}$ (which corresponds to the spatial inverse metric) 
 and by a nowhere vanishing vector $n_\mu$ (defining the local time direction),
  with the hortogonality condition
 \beq
 \label{nort}
 n_\mu h^{\mu \nu}=0\ .
 \eeq 
A velocity field $v^\mu$ is also introduced, 
with the condition 
\beq
\label{nort2}
n_\mu v^\mu =1 \, .
\eeq
 Given 
$(h^{\mu \nu}, n_\mu, v^\nu)$,  one can then uniquely define the spatial metric 
$h_{\mu \nu}$, with:
\beq
\label{nort3}
h^{\mu \rho}h_{\rho \nu}=\delta^\mu_\nu -v^\mu n_\nu\equiv P^\mu_\nu \, , \qquad
h_{\mu \a} v^\a =0 \, ,
 \eeq
 where $P^\mu_\nu$ is the projector onto spatial directions.
We introduce also a non-dynamical gauge field $A_\mu$ 
 as a source for the particle number symmetry\footnote{The presence of the vector field is related to 
 the arbitrariness $v^\mu \to v^\mu + h^{\mu \nu}A_\nu$  one has in defining the velocity field.}.
In term of the $2+1$ dimensional quantities, the extra 
dimensional metric used in the null reduction is:
\beq
\label{null-red}
G_{MN}= \begin{pmatrix}
0  & n_{\nu}  \\
n_{\mu}  & n_{\mu} A_{\nu} + n_{\nu} A_{\mu} + h_{\mu\nu}
\end{pmatrix} \, ,  \qquad
G^{MN}= \begin{pmatrix}
A^2 - 2 v \cdot A  & v^{\nu}-h^{\nu\sigma} A_{\sigma}  \\
v^{\mu}-h^{\mu\sigma} A_{\sigma}  & h^{\mu\nu}
\end{pmatrix} \, ,
\eeq
where $A^2 - 2 v \cdot A = h^{\mu \nu} A_\mu A_\nu - 2 v^\mu A_\mu$. 
The determinant of the metric  is then
\beq
\sqrt{g}= \sqrt{- \det G_{AB}}=
\sqrt{{\rm det}(h_{\mu \nu}+ n_\mu n_\nu)} \ .
\eeq
The symmetries of the Newton-Cartan theory include, besides diffeomorphisms and local U(1) gauge invariance, a local version of Galilean boosts, namely the Milne boosts.
Transformation properties of fields under Weyl and  Milne boosts  are reported in Appendix \ref{appeweyl}.

To deal with spinors, it is necessary to introduce an orthonormal frame 
field which relates the metric in the curved spacetime with the flat tangent space. The flat tangent space metric is:
\beq
\label{flat}
G_{AB}= G^{AB} = \begin{pmatrix}
0  & 1 & 0 & 0  \\
1  & 0 & 0 & 0  \\
0  & 0 & 1 & 0  \\
0  & 0 & 0 & 1  \\
\end{pmatrix} \, , 
\eeq
and the frame fields are defined by:
\bea
& G_{MN}= e^{A}_{\,\,\, M} G_{AB} e^{B}_{\,\,\, N}  \, ,  
\qquad  G_{AB}= e^{M}_{\,\,\, A} G_{MN} e^{N}_{\,\,\, B}  \, ,  
\nl
&  e^{A}_{\,\,\, M} e^{M}_{\,\,\, B} = \delta^{A}_{\,\,\, B}  \, , \qquad
  e^{M}_{\,\,\, A} e^{A}_{\,\,\, N} = \delta^{M}_{\,\,\, N}  \, .
\eea
The spin connection associated to the vielbein is
\bea
\omega_{MAB} &=& \frac{1}{2} \left[ e^{N}_{\,\,\,A} \left( \p_M  e_{NB} - \p_N   e_{MB} \right) 
-  e^{N}_{\,\,\,B}  \left( \p_M  e_{NA} - \p_N   e_{MA} \right)  \right. \nl
& & \left. 
- e^{N}_{\,\,\,A} e^{P}_{\,\,\,B}  \left( \p_N  e_{PC} - \p_P   e_{NC} \right)   e^{C}_{\,\,\,M} \right] \, .
\eea

The explicit form of the fierbein, dreibein and their inverses as well as conventions on gamma matrices are summarized in Appendix \ref{appegamma}.


\subsection{Sources and conserved currents}

Let us consider a generic nonrelativistic matter field $\phi$ coupled to a NC background (collectively denoted by $g$).
The vacuum functional $W[g]$ is defined as the quantum average of the action over the matter fields, namely as the path integral
\beq
e^{i W[g]}= \int {\cal D} \phi^*  \, {\cal D} \phi \,  e^{i S[\phi , \, \phi^* \!, \, g]} \ .
\eeq
The NC fields act as sources in the vacuum functional, so that it 
is possible to generate all the Ward identities by functional differentiation of $W[g]$ with respect to the NC sources. 
However, due to the constraints  (\ref{nort}) - (\ref{nort3}) relating the NC entrees,
arbitrary variations on background fields are not allowed,
and one must first identify the independent variations, 
{\it e.g.}~\cite{Geracie:2014nka}.
These  can be parameterized in terms
of an arbitrary $\delta n_\mu$, a transverse
 perturbation $\delta u^\mu$ with
 $\delta u^\mu n_\mu =0$ and a transverse metric
 perturbation $\delta \th^{\a \b} n_\b=0$.
Then, the variation of the NC metric in terms of the independent variations can be written as
 \beq
 \delta n_\mu \, , \qquad
 \delta v^\mu = - v^\mu v^\a \delta n_\a + \delta u^\mu \, , \qquad
 \delta h^{\mu \nu} =-v^{\mu} \delta n^{\nu}-\delta n^\mu v^\nu
 -\delta \th^{\mu \nu} \, .
 \label{varii}
 \eeq
 Consequently, the NC metric nearby the flat limit (\ref{flat}) gives
 \bea
 n_\mu  &=&  (1+\delta n_0 , \delta n_i) \, , \qquad
 v^\mu =(1-\delta n_0, \delta u_i) \, , \qquad \delta \th^{0 i}=0 \, ,
 \nl
  h_{\mu \nu} &=&
  \left(
\begin{array}{cc}
 0& -\delta u_i \\ -\delta u_i  &  \delta_{ij} + \delta \th_{ij} \\
  \end{array}\right) \, , \qquad
h^{\mu \nu}=
  \left(
\begin{array}{cc}
 0& -\delta n_i \\ -\delta n_i  &  \delta_{ij} - \delta \th_{ij} \\
  \end{array}\right) \, .
\eea
The null reduction metric is
 \bea
G_{A B} & = &
  \left(
\begin{array}{ccc}
0 & 1 + \delta n_0 &  \delta n_i \\
1+\delta n_0 & 2 \delta A_0 & \delta A_i -\delta u_i \\
\delta n_i  & \delta A_i -\delta u_i &\delta_{ij} + \delta \th_{ij} \\
  \end{array}\right) \, ,
  \nl
  G^{A B} & = &
  \left(
\begin{array}{ccc}
-2 A_0 & 1- \delta n_0 & -\delta A_i+ \delta u_i \\
1-\delta n_0 &  0& -\delta n_i \\
-\delta A_i+ \delta u_i  &  -\delta n_i&\delta_{ij} - \delta \th_{ij} \\
  \end{array}\right) \, .
  \label{perturba}
\eea

We can use these sources to define the currents of the energy momentum tensor multiplets: 
under the above infinitesimal variations, 
the vacuum functional varies according to
\beq
\delta W = \int d^d x\sqrt{-g} \le
\frac{1}{2} T_{i j} \delta \th_{i j} + j^\mu \delta A_\mu
-\epsilon^\mu \delta n_\mu - p_i \delta u_i
\ri \, .
\eeq
Here $p_i$ is the momentum density,
$T_{i j}$ is the spatial stress tensor, $j^\mu=(j^0,j^i)$ contains
the $U(1)$ number density and current, and $\epsilon^\mu=(\epsilon^0,\epsilon^i)$
 is the energy density and current. 
 Here $\epsilon^\mu$ includes also the contribution coming from
the ``chemical potential" $A_0$, see e.g. eq.~(\ref{actionscalar}).
 Note that in eq.~(\ref{perturba}),
 the variations $\delta A_i$ and $\delta u_i$ do not appear independently 
 but always in the combination $\delta A_i - \delta u_i$.
 As a consequence of this fact, the $U(1)$ number particle current 
 is always proportional to  the momentum density, 
 as it should be in a nonrelativistic theory.

Ward identities  in the flat limit can be easily obtained in the usual way: consider a symmetry of the classical action,  specify  the 
corresponding infinitesimal variation of the NC metric, and impose invariance of the functional, $\delta W =0$.
Then, associated to particle number conservation there is conservation of the $U(1)$ current
\beq
\langle \p_\mu  j^\mu \rangle =0  \ .
\eeq
Associated to diffeomorphism invariance there are the conservation of the spatial stress tensor and the energy current conservation
\beq
\label{diffcurrent}
\langle \p_t p^j  + \p_i T^{ij }\rangle =0 \ , \quad \langle \p_\mu  \epsilon^\mu\rangle=0 \, .
\eeq 
Finally, local Weyl transformation entails the Ward identity associated to the conservation of the scale current,  
which is found to be\footnote{Strictly speaking, the scale current has an additional term proportional to the scaling dimension $\Delta$ of the matter field. However, such term is a total derivative and can always be reabsorbed by a current redefinition. }
\beq
\label{defscalecurrent}
J_S^0 = p_i x^i - 2t \epsilon^0 \, , \quad
J_S^i =  x^j T^i_{\,\,\, j} -2t \epsilon^i  \, ,  \quad \langle \p_\mu  J_S^\mu \rangle =0\ .
\eeq
By expanding explicitly the scale Ward identity we have
\beq
\langle \p_\mu J_S^\mu \rangle=
\langle T^i_{\,\,\, i} - 2 \epsilon^0\rangle - 2t \langle \p_\mu  \epsilon^\mu\rangle  + 
x^j \langle  \p_t  p_j + \p_i  T^{i }_{\ j}\rangle  = 0 \, .
\label{Classical scale nonrel}
\eeq
Equation (\ref{Classical scale nonrel}) is interesting, as it reveals the relations intertwining 
 between tracelessness of the energy-momentum tensor,
conservation of the energy momentum tensor and scale conservation. A quantum violation of the scale symmetry manifests 
as a non conservation of the scale current $J_S^\mu$ which, in turn,  is equivalent to  a violation of the tracelessness condition
 $\langle T^i_{\,\,\, i} - 2 \epsilon^0\rangle =0$ only if the energy-momentum tensor does not have a
 diffeomorphism anomaly, {\it i.e.} only if the conditions (\ref{diffcurrent}) are satisfied. On the contrary, if the energy momentum tensor 
 is not conserved at the quantum level, not only the trace anomaly, but also the diffeomorphism anomaly  contribute to the scale anomaly.



\subsection{Flat spacetime with $U(1)$ gauge field}

We will compute the trace anomaly
for a flat background where only the non-dynamical $U(1)$ gauge potential
 is switched on:
\beq
n_{\mu} = (1, \mathbf{0}) \, , \qquad
v^{\mu} = (1, \mathbf{0}) \, , \qquad
h_{ij} = \delta_{ij} \, , \qquad
A_{\mu} = (A_0 (t,x^i), A_i (t,x^i)) \, ,
\eeq
which corresponds to the extra dimensional metric
\beq
G_{MN} = \begin{pmatrix}
0 & 1 & 0 & 0 \\
1 & 2A_0 & A_1 & A_2 \\
0 & A_1 & 1 & 0 \\
0 & A_2 & 0 & 1 \\
\end{pmatrix} \, , \qquad
G^{MN} = \begin{pmatrix}
-2 A_0+ A_i A_i & 1 & -A_1 & -A_2 \\
1 & 0 & 0 & 0 \\
-A_1 & 0 & 1 & 0 \\
-A_2 & 0 & 0 & 1 \\
\end{pmatrix} \, .
\eeq
The  vielbein is:
\beq
e^A_{\,\,\, M} = \begin{pmatrix}
1 & A_0 & A_1 & A_2 \\
0 & 1 & 0 & 0 \\
0 & 0 & 1 & 0 \\
0 & 0 & 0 & 1 \\
\end{pmatrix} \, , \qquad
e^M_{\,\,\, A} = \begin{pmatrix}
1 & -A_0 & -A_1 & -A_2 \\
0 & 1 & 0 & 0 \\
0 & 0 & 1 & 0 \\
0 & 0 & 0 & 1 \\
\end{pmatrix}\, .
\eeq
Starting from these data, we can compute the non-vanishing components of the spin connection
\beq
\omega_{++i} = - F_{0i} \, ,  \qquad
\omega_{+ij} = - \frac12 F_{ij} \, , \qquad \omega_{i+ j} = - \frac12 F_{ij} \, ,
\eeq
and the non-vanishing components of the Christoffel symbol\footnote{In the following, the presence of the subscript $A$ denotes that the corresponding quantity is Milne boost invariant}:
\beq
\Gamma^{-}_{\,\,\, \mu \nu} =\frac12 (v_A)^{\sigma} (Q_A)_{\mu\nu\sigma} \, , \qquad  \Gamma^{\rho}_{\,\,\, \mu\nu} =\frac12 h^{\rho \sigma} (Q_A)_{\mu\nu\sigma} \, ,
\eeq
where 
\bea
F_{\mu\nu} &=& \p_{\mu} A_{\nu} - \p_{\nu} A_{\mu} \, , \cr
(v_A)^\mu &=& v^\mu - h^{\mu \nu} A_\nu \, , \cr
(h_A)_{\mu \nu}&=&h_{\mu \nu} +A_\mu n_\nu + A_\nu n_\mu \, ,\cr
(Q_A)_{\mu \nu \sigma} &=& \p_\mu (h_A)_{\nu \s} +\p_\nu (h_A)_{\mu \s} - \p_\s (h_A)_{\mu \nu} \, .
\eea

\section{Trace anomaly}
\label{tratra}

The action for a nonrelativistic boson is
\beq
S=  \int d^{3}x \sqrt{g} \,   \left[ im v^{\mu} \phi^{\dagger}  D_\mu   \phi -im v^{\mu} ( D_\mu \phi )^{\dagger}  \phi 
 - h^{\mu\nu} ( D_\mu \phi )^{\dagger} D_\nu  \phi - \xi R \phi^{\dagger} \phi \right]\, ,
\eeq
where $D_\mu =  \p_{\mu} - im A_{\mu}$.
Specializing to a flat background and performing 
an integration by parts, we find
\beq
\label{actionscalar}
S = \int d^{3} x \, \left[ 2 im \phi^{\dagger} \p_t \phi
 +    \phi^{\dagger} \p_i^2 \phi - 2 im A_i \phi^{\dagger} \p_i \phi 
 + \left( 2 m^2 A_0  - m^2 A_i A_i  - i m \p_i A_i \right) 
 \phi^{\dagger} \phi  \right] \, .
\eeq
 Note that $A_0$ plays the role of a grand canonical chemical potential
coupled to the particle number $J_0=2m^2 \phi^\dagger \phi$. 
As a consequence, $\epsilon^0= E +   A_0 J^0$,
where $E$ is the particle energy density.

\subsection{The unperturbed case}
We shall compute  the trace anomaly 
using the Heat Kernel (HK) method in imaginary time space. 
The following substitutions are used \cite{Solodukhin:2009sk}
\beq
t \rightarrow -i t_E \, , \qquad \p_t \rightarrow i \p_{t_E} \, , \qquad m \rightarrow im_E \, .
\label{wick-rotation}
\eeq 
For a generic operator ${\hat {\cal O}}_E$, the HK operator of  ${\hat {\cal O}}_E$ is defined as
\beq
\hat{K}_{{\hat{{\cal O}}_E}}(s)=\exp (s {\hat{{\cal O}}_E} ) \, .
\eeq
The following matrix elements are introduced
\beq
K_{{\hat{{\cal O}}_E}}(s,x,t,x',t') = \langle x  t | \hat{K}_{{\hat{{\cal O}}_E}}(s)| x' t' \rangle \, ,
\eeq
and we denote by $\tilde{K}_{{\hat{{\cal O}}_E}}$ the diagonal matrix elements
\beq
\label{tilde}
\tilde{K}_{{\hat{{\cal O}}_E}}(s,x,t) = \langle x  t | \hat{K}_{{\hat{{\cal O}}_E}}(s)| x t \rangle \, .
\eeq
The HK is an efficient method to compute the one-loop effective action;
in the unpertubed flat nonrelativistic case with $A_\mu=0$,
the free Schr\"odinger operator  $\triangle$ entering the action  is given by 
\beq
 \triangle = \left( - 2 i m \p_t + \p_i^2 \right) =
\left( - 2  m \sqrt{- \p_t^2} + \p_i^2 \right) \, .
\eeq
The HK matrix elements $K_\triangle$ have been computed in \cite{Auzzi:2016lxb},
\beq
\label{heatfree}
K_\triangle(s)=\langle x  t | e^{s \triangle} | x' t' \rangle=
\frac{1}{2 \pi} \, \frac{ms}{m^2 s^2 +\frac{(t-t')^2}{4}}  \,
\frac{1}{(4 \pi s) } \exp \le -\frac{(x-x')^2}{4 s} \ri \, .
\eeq

\subsection{The perturbative expansion}

In curved space one introduces the following scalar product
in coordinate representation:
\beq
\langle xt | x' t' \rangle_g = \frac{\delta(x-x')\delta(t-t')}{\sqrt{g}} \, . 
\eeq
In a generic background, it is convenient to expand the complete Schr\"odinger operator $\hat \triangle$
as the sum of its free part $\triangle$ plus a perturbation $ \hat V$.
We can then evaluate the 
 HK  as a perturbative expansion around (\ref{heatfree}). 
The diagonal elements of the HK operator can be expanded 
in powers of $s$ as:
\beq
\tilde{K}_{\hat{ \triangle}}(s)=
{\rm{Tr}}
\langle x  t | e^{s \hat{ \triangle}} | x t \rangle_g=
\frac{1}{s^{2}} 
\le a_0(\hat{ \triangle}) +a_2(\hat{ \triangle}) s + a_4(\hat{ \triangle}) 
s^2 + \dots \ri \, .
\eeq
This expansion provides the definition of
 the  De~Witt-Seeley-Gilkey coefficients $a_{2k}(\hat{ \triangle})$.
For a  nonrelativistic $2+1$ dimensional theory, the trace anomaly 
is proportional to the $a_4$ coefficient \cite{Auzzi:2016lxb}.

In general, it is convenient to work in a 
quantum mechanical space with flat inner product  
\beq
\langle xt | x' t' \rangle = \delta(x-x')\delta(t-t') \, .
\eeq
Consequently, for any operator $\hat{\cal O}$ we can define the operator $ \hat{M}_{\hat{\cal O}} $ such that 
\beq
\langle xt | \hat{\cal O} | x't' \rangle_g = \langle xt | \hat{M}_{\hat{\cal O}} |x' t' \rangle \, .
\eeq
The ``effective'' operator $ \hat{M}_{\hat{\cal O}}$
keeps track of the metric in the inner product.
In this way we can  expand the diagonal elements  of 
the HK as
\beq
\label{k-trace}
\tilde{K}_{\hat{M}} (s) = {\rm{Tr}} \, \langle xt | e ^{s \hat{M}} | x t \rangle =
\frac{1}{s^{2}} \left[ a_0 ( \hat{M}) + s \, a_2 (\hat{M}) + s^2 a_4 (\hat{M}) + \dots \right] \equiv
\sqrt{g} \tilde{K}_{\hat{\bigtriangleup}} (s)  \, .
\eeq
In our flat case, $\sqrt{g}=1$ and so
$\hat{\cal O}= \hat{M}_{\hat{\cal O}} $.


We can parameterize  the perturbation from the flat contribution as:
\beq
\begin{aligned}
& \langle x t | \hat{M} | x' t' \rangle = \langle x t | \bigtriangleup + \hat{V} | x' t' \rangle 
=  \langle x t | \bigtriangleup + P(x) \delta (x-x') \delta (t-t') \\
 &  +
S(x) \sqrt{-\p_t^2} \delta (x-x') \delta (t-t')  + Q_i (x) \p_i \delta (x-x') \delta (t-t') | x' t' \rangle \, . 
\label{decomposition flat operator}
\end{aligned}
\eeq
The perturbative calculation starts by considering the expansion
\beq
K_{\hat{M}} (s) = \exp \le  s (\bigtriangleup + \hat{V}) \ri = \sum_{n=0}^{\infty} K_n (s) \, ,
\eeq
where the single terms entering the series are obtained via a Dyson recursive procedure:
\beq
K_n (s) = \int_0^s ds_n \int_0^{s_n} ds_{n-1} \dots \int_0^{s_2} ds_1 e^{(s-s_n)\bigtriangleup} \hat{V} \dots e^{(s_2-s_1)\bigtriangleup} \hat{V} e^{s_1 \bigtriangleup} \, .
\eeq
The terms $ K_n (s) $ consists of insertions of \emph{n} operators among the set
\beq
\lbrace P(x), S(x), Q_i (x)  \rbrace 
\eeq
and then the computation is performed when we find the value of $ K_n (s) $ for all the non-vanishing combinations of terms in the previous set.

The imaginary time rotation of the gauge field gives\footnote{The unconventional redefinition of 
the gauge field  in the imaginary time formalism is required by consistency with  $[D_\mu, D_\nu] = - i m F_{\mu \nu}$ and the prescription $m \rightarrow i m$. The imaginary mass is required in order to get a positive definite euclidean action.}:
\beq
A_0 \rightarrow A_0 \, \qquad \, A_i \rightarrow -i A_i \, ,
\eeq
and the imaginary time action reads:
\beq
S_E   = -  \int d^{3} x \,  \phi^{\dagger} \left[ \bigtriangleup - 2 i m A_i  \p_i   - 2  m^2 A_0 - m^2 A_i A_i  - i m (\p_i A_i)   \right] \phi \, .
\eeq
We can immediately identify
\beq
\label{functions}
S(t,x^i) = 0 \, , \quad
P(t,x^i) = -2  m^2 A_0 - m^2 A_i A_i  - i  m (\p_i A_i)  \, , \quad
Q_i (t,x^i) = - 2 im A_i \, ,
\eeq
so that all the insertions containing at least one operator $ S(t,x^i) $ vanish.
Therefore, at the first order ($n=1$), there are just two terms, denoted by ${K}_{1P}$ and ${K}_{1Q_i}$. At the second order (double insertion, $n=2$) we have four possible insertions, ${K}_{2PP}$, ${K}_{2PQ_j}$, ${K}_{2Q_iP}$, ${K}_{2Q_i Q_j}$. 

For time independent backgrounds, the calculation of the coefficients can be found in \cite{Auzzi:2016lxb}-\cite{Auzzi:2017jry}
with the exception of the ${K}_{1Q_i}$ term, which was not needed neither for the scalar nor for the fermion anomaly.
Here in Appendix \ref{AppeA} we calculate such a term. In Appendix \ref{AppeB}, instead, we provide all the generalizations needed for all the terms 
when the background is time dependent.
For the single insertion, we get 
\beq
\tilde{K}_{1P} = \frac{1}{8 m \pi^2 s^2} \mathrm{Tr} \left( s P + \frac{1}{6} s^2 \p_x^2 P + \dots \right) \, ,
\label{inizio 1 inserzione}
\eeq
\beq
\tilde{K}_{1Q_i} = \frac{1}{8 m \pi^2 s^2} \mathrm{Tr} \left( - \frac{s}{2} \p_i Q_i - \frac{s^2}{12} \p_i \p^2 Q_i  + \dots \right) \, ,
\label{fine 1 inserzione}
\eeq
where $ \Tr $ denotes a trace over indices such as internal or spinorial ones (in the scalar case the trace is redundant).  Substituting eq.(\ref{functions}) in (\ref{inizio 1 inserzione}) and (\ref{fine 1 inserzione}),
and matching the powers of $s$ in eq. (\ref{k-trace}) one can get the contribution to the $a_2$ and $a_4$ coefficients 
 coming from the single insertion. 

Concerning the contribution coming from the double insertion,
from Appendix \ref{AppeB} one gets the general formulae 
\beq
\tilde{K}_{2PP} = \frac{1}{8 m \pi^2 s^2} \mathrm{Tr} \left( \frac{s^2}{2} P(x)^2 + \dots \right)  \, ,
\label{inizio inserzioni doppie}
\eeq
\beq
\tilde{K}_{2 a_j a_i } = \frac{1}{8 m \pi^2 s^2} 
\textrm{Tr}  \left[ -  \frac{s}{4}  Q_i Q_i   - \frac{s^2}{12} Q_i  (\p_{i} \p_{j} Q_j ) + \frac{s^2}{12}  (\p_{i} \p_{j} Q_i) Q_j   - \frac{s^2}{24} (\p_{j} Q_i) (\p_i Q_j) \right.
\eeq
\[  
\left.  
+  \frac{s^2}{8} (\p_{i} Q_i) (\p_{j} Q_j)   -  \frac{s^2}{12} Q_i (\p^2 Q_i) 
 -  \frac{s^2}{24} (\p_i Q_j)^2 
+  \dots \right] \, ,
\]
\beq
\tilde{K}_{2 a_i P}  =  \frac{1}{8 m \pi^2 s^2} \mathrm{Tr} \left( - \frac{s^2}{3} P (\p_i Q_i)  - \frac{s^2}{6} (\p_i P) Q_i  + \dots \right)  \, ,
\eeq
\beq
\tilde{K}_{2  P a_i}  =  \frac{1}{8 m \pi^2 s^2}\mathrm{Tr} \left(   \frac{s^2}{6} Q_i (\p_i P) - \frac{s^2}{6} (\p_i Q_i) P + \dots \right)  \, .
\label{fine 2 inserzioni}
\eeq
With the above formulas, in a similar way, one gets the contributions to $a_2$ and $a_4$ coming from the double insertions. Summing all together, and extracting the null power of $s$ in (\ref{k-trace}), we easily get the $a_4$ coefficient up to the second order in the fields, and therefore the trace anomaly
\beq
\label{a4scalar}
a_4 = \langle T^i_{\ i} - 2 \epsilon^0 \rangle = - \frac{m}{8 \pi^2} \left( \frac13 \p^2 A_0 + \frac16  B^2 - {2 m^2} A_0^2 + \mathcal{O} (A_{\mu}^3)\right)\equiv {\cal A} \, ,
\eeq
where $B= F_{12}$. Equation (\ref{a4scalar}) deserves few comments. First of all, the anomaly 
breaks both Milne boost  and gauge invariance.  Due to 
 the intimate relationship  intertwining the two symmetries\footnote{In the Bargmann algebra, the commutator of the momentum and a boost is the particle number generator.}, it is not surprising that breaking one of them does entail the breaking of the other. 
  In addition, note that in (\ref{a4scalar}) the  
$\p^2A_0$ term does not serve to rebuild a divergence of the electric field, as it would be
if the result were gauge invariant.
 Rather,  in (\ref{a4scalar}), $A_0$ should be considered as $v^\mu A_\mu$, otherwise the first two terms  would not have the correct Weyl weight (see Appendix \ref{appeweyl}).
 
Concerning the $\p^2 (v^\mu A_\mu)$ term  in 
eq.~(\ref{a4scalar}), it can be reabsorbed by a local counterterm 
in the vacuum functional  $W$ proportional to $R \, v^\mu A_\mu$. This is not possible for the  $(v^\mu A_\mu)^2$
term, that is therefore a (type-B) genuine anomaly.

Both in the free scalar and free fermion examples, the field $A_0$ plays the role of an external chemical potential
for the particle number $J_0$; in the multiple species case, $J_0$ plays the role of mass density.
Moreover, studying geodesics in a NC background, one sees that $A_0$ can also be identified 
 as the Newtonian gravitational potential. On physical ground one would expect mass conservation
in an external gravitational field. On the other hand, the breaking of gauge invariance
in eq. (\ref{a4scalar})  may hint a violation of the conservation of
the $U(1)$ current; if this would be the case, this would be puzzling because it would 
not be consistent with the physical intuition. This point is beyond the purpose of the present paper
and deserves further investigation.

\subsection{The fermion}

The  Dirac operator is expressed as
\beq
\slashed D= \gamma^{M} D_{M} = \gamma^{A} e^{M}_{\,\,\, A} D_{M}  \, ,
\eeq
Conventions on gamma matrices with lightcone indices are summarized  in Appendix \ref{appegamma}
and are the same  used in \cite{Auzzi:2017jry}.
The covariant derivative acting on fermions is
\beq
D_{M} \Psi =\left( \p_{M} + \frac{1}{4} \omega_{MAB} \gamma^{AB} \right) \Psi = \left( \p_{M} + \frac{1}{8} \omega_{MAB} [\gamma^{A}, \gamma^{B}] \right) \Psi \, ,
\eeq
$\omega_{MAB}$ being the spin connection.
We can write the nonrelativistic fermion action in $2+1$-dimensions
from the null reduction of the $3+1$-dimensional Dirac action:
\beq
S= \int d^4 x \sqrt{g} \, i \bar{\Psi} \slashed D \Psi \, ,
\label{didirac}
\eeq
using the following profile for the fermion along the extra dimension:
\beq
\Psi(x^M)=\psi(x^{\mu}) e^{imx^{-}} \, .
\eeq

In the fermionic case, the imaginary-time Dirac operator $\slashed D$ is not elliptic.
In order to avoid this difficulty, the squared Dirac operator $\slashed D^2$
is used to compute the vacuum functional:
\beq
\label{eqdet2}
i W= \frac12 \log \det (\slashed D^2) \, .
\eeq
This trick is used both in the relativistic, see {\it e.g.} \cite{Christensen:1978md}, 
and nonrelativistic case \cite{Auzzi:2017jry}.
Specializing to a flat background geometry, {\it i.e.} $\sqrt{g} = 1$, $R=0$,
and going to imaginary time, we find:
\beq
\slashed D^2_E \Psi = \bigtriangleup \Psi - 2  m^2 A_0 \Psi -  m^2 A_k A_k \Psi 
- i  m (\p_i A_i) \Psi - 2 i m A_i (\p_i \Psi) +
\eeq
\[
- m F_{i0} \gamma^{+i} \Psi  - \frac{1}{4} i  m F_{ij} \gamma^{ij} \Psi + \frac12  m A_i F_{ij} \gamma^{+j} \Psi + \frac12 i  F_{ij} \gamma^{+j}(\p_i \Psi) + \frac{1}{4} i  (\p_i F_{ij}) \gamma^{+j} \Psi \, .
\]
According to eq.(\ref{decomposition flat operator}) we can identify 
\beq
S(t,x^i) = 0 \, , \qquad
Q_i (t,x^i) = \le - 2 i m A_i \ri \mathbf{1} + \frac12 i F_{ij} \gamma^{+j} \, ,
\eeq
\bea
P(t,x^i)&=& \left[ -2  m^2 A_0 - m^2 A_k A_k - i m (\p_i A_i) \right] \mathbf{1} 
\nonumber
\\
&&- m F_{i0} \gamma^{+i} -
 \frac{1}{4} i m F_{ij} \gamma^{ij}  + \frac12  m A_i F_{ij} \gamma^{+j} + \frac{1}{4} i (\p_i F_{ij}) \gamma^{+j} \, ,
\eea
where the Dirac matrices read:
\beq
\gamma^{+1} = \begin{pmatrix}
0 & \sqrt{2} & 0 & 0 \\
0 & 0 & 0 & 0 \\
0 & 0 & 0 & 0 \\
0 & 0 & - \sqrt{2} & 0 \\
\end{pmatrix} \, , \ 
\gamma^{+2} = \begin{pmatrix}
0 & -\sqrt{2} i & 0 & 0 \\
0 & 0 & 0 & 0 \\
0 & 0 & 0 & 0 \\
0 & 0 & - \sqrt{2} i & 0 \\
\end{pmatrix} \, , \ 
\gamma^{12} = \begin{pmatrix}
-i & 0 & 0 & 0 \\
0 & i & 0 & 0 \\
0 & 0 & -i & 0 \\
0 & 0 & 0 & i \\
\end{pmatrix} \, .
\eeq

We can now use
eqs. (\ref{inizio 1 inserzione})-(\ref{fine 1 inserzione})
and eqs. (\ref{inizio inserzioni doppie})-(\ref{fine 2 inserzioni})
to evaluate the single and double insertion contributions.

Summing both the first and the second order terms in 
the external background fields, we find
\beq
a_4 (\slashed D^2_E) 
= - \frac{m}{48 \pi^2} B^2 - \frac{m}{6 \pi^2} \p^2 A_0 + \frac{m^3}{\pi^2} A_0^2  + \mathcal{O}(A_{\mu}^3) \, .
\eeq
The trace of the stress-energy tensor is finally given by
\beq
\langle T^i_i - 2 T^0_0 \rangle = - \frac12 a_4 (\slashed D^2_E)   = \frac{m}{12 \pi^2} \p^2 A_0 - \frac{m^3}{2 \pi^2} A_0^2 + \frac{m}{96 \pi^2} B^2   + \mathcal{O}(A_{\mu}^3)  \, . 
\eeq
The structure is the same of the bosonic case.


\section{Diffeomorphism anomaly}
\label{diffe}

The previous calculation of the Weyl anomaly, which related the trace of the energy-momentum tensor with the $a_4 $ coefficient, relies on the $ \zeta $ function regularization.
This method is described in \cite{Vassilevich:2003xt} and here we sketch the derivation\footnote{ Reference \cite{Vassilevich:2003xt} is very exhaustive but also pretty long. For the benefit of the reader, we recall that Sect. 2.2 deals with the spectral functions relevant to this section, whereas the part relevant to the conformal anomaly can be found in Sect. 7.1.}. For an 
 operator $ \mathcal{D}$ defining a classical action, 
 the regularized vacuum functional is defined by 
 \beq
W^{\mathrm{reg}}(s)=-\frac12 \tilde \mu^{2s} \int_0^\infty\frac{dt}{t^{1-s}} \tilde K_\mathcal{D}(t)
=-\frac12 \tilde \mu^{2s} \Gamma(s) \zeta(s,\mathcal{D})\ ,
 \eeq
where the regulator is $s$, and $\tilde \mu$ is the mass parameter that any regularization procedure entails.
The last equality gives the relation between the HK spectral function and the zeta function of an operator
\beq
\label{zetatraccia}
\zeta (s, \mathcal{D}) = \frac{1}{\Gamma(s)} \int_0^\infty\frac{dt}{t^{1-s}} \tilde K_\mathcal{D}(t)
= \Tr (\mathcal{D}^{-s}) \, .
\eeq
The physical limit of the regularized vacuum functional is attained for $s\to 0$ that, 
due to the presence of the $\Gamma$ function, develops an UV  singularity that needs to be subtracted. The renormalized vacuum functional is the $s\to 0$ limit of the subtracted vacuum functional, leading to 
\beq
W^{\mathrm{ren}} = - \frac12 \zeta' (0, \mathcal{D}) - \frac12 \log \mu^2 \zeta(0, \mathcal{D}) \, ,
\eeq
where $ \mu^2 = e^{-\gamma_E} \tilde \mu^2 $ is the renormalization scale and $\gamma_E$ the Euler Mascheroni constant.

Next, we need to see how the renormalized vacuum functional varies under a variation $\delta \mathcal{D}$.  This is completely determined by the variation of the $\zeta (s, \mathcal{D})$ function\footnote{Note that we need to compute the variation for $s\ne 0$ and eventually perform the $s\to 0$ limit.}, 
\beq
\delta \zeta (s, \mathcal{D}) = -s \Tr \le (\delta \mathcal{D}) \mathcal{D}^{-s-1} \ri \, .
\eeq
To compute the diffeomorphism anomaly, we need the variation
$ \delta \mathcal{D} $  under diffeomorphisms.
In the scalar case, after integration by parts, the 
imaginary-time action can be put in the form
\beq
S_E =  \int d^3 x \sqrt{g} \,  \phi^{\dagger} \mathcal{D} \phi \, ,
\eeq
with
\beq
\mathcal{D}\phi  = i m v^{\mu} D_{\mu} \phi + \frac{i m }{\sqrt{g}} D_{\mu} \le \sqrt{g} v^{\mu} \phi \ri - \frac{1}{\sqrt{g}} D_{\mu} \le \sqrt{g} h^{\mu\nu} D_{\nu} \phi\ri \, .
\eeq
We will only consider the variation under diffeomorphisms of this operator specializing  to a flat background with  a non-vanishing gauge field. The scalar operator specialized to  this background is
\beq
\mathcal{D}_0 = 2 i m \p_0 - \p_i^2 +2 m^2 A_0 + m^2 A_i A_i + 2 i m A_i \p_i + i m (\p_i A_i) \, ,
\eeq
and it transforms under diffeomorphisms as:
\bea
\delta \mathcal{D}_0   &=& - 2 i m (\p_0 \varepsilon^{\mu}) \p_{\mu} + 2 (\p_i \varepsilon^{\mu}) \p_i \p_{\mu} + (\p_i^2 \varepsilon^{\mu}) \p_{\mu} + 2 m^2 \varepsilon^{\mu} (\p_{\mu} A_0)
\nonumber  \\
& & + 2 i m \varepsilon^{\mu} (\p_{\mu} A_i) \p_i - 2 i m A_i (\p_i \varepsilon^{\mu}) \p_{\mu} + i m \varepsilon^{\mu} (\p_i \p_{\mu}A_i) + 2 m^2 A_i \varepsilon^{\mu} (\p_{\mu} A_i) \, .
\label{variazione sotto diffeo operatore}
\eea

Now comes an important point, that permits to understand which terms in eq. (\ref{variazione sotto diffeo operatore}) do indeed
contribute to the anomaly. The $\zeta$ function (\ref{zetatraccia}) is a trace. As such, due to the cyclicity properties,  is invariant under the  similarity transformation 
\beq
\label{dtilde}
\zeta (s, \tilde{\mathcal{D}}) = \zeta (s, \mathcal{D}) \, \qquad \text{ if} \quad\tilde{\mathcal{D}} = e^{\hat{O}} \mathcal{D} e^{-\hat{O}} \, .
\eeq
Strictly speaking, this means that $\mathcal{D}$ and $\mathcal{\tilde D}$ have the same functional determinant.
If we consider the redefinition (\ref{dtilde}) with 
\beq
\hat{O} = \alpha \xi^{\mu} \p_{\mu} \, ,
\eeq
with $ \alpha $  a real coefficient and $ \xi^{\mu} $ transforming under diffeomorphisms as
$\delta \xi^{\mu} = \varepsilon^{\mu} $,
we find that
\begin{align} \nonumber
\tilde{\mathcal{D}_0} = e^{\alpha \xi^{\mu} \p_{\mu}} \mathcal{D}_0 e^{- \alpha \xi^{\nu} \p_{\nu}} & = \mathcal{D}_0 - 2 i m \alpha (\p_0 \xi^{\mu}) \p_{\mu} + 2 \alpha (\p_i \xi^{\mu}) \p_i \p_{\mu} + \alpha (\p_i^2 \xi^{\mu}) \p_{\mu} \\
& + 2 m^2 \alpha \xi^{\mu} (\p_{\mu} A_0) + 2 m^2 \alpha A_i \xi^{\mu} (\p_{\mu} A_i) + 2 i m \alpha \xi^{\mu} (\p_{\mu} A_i) \p_i \\
\nonumber
& + i m \alpha \xi^{\mu} (\p_{\mu} \p_i A_i) - 2 i m \alpha A_i (\p_i \xi^{\mu}) \p_{\mu} + \mathcal{O} (\xi^2) \, .
\end{align}
Using eq. (\ref{variazione sotto diffeo operatore}) and setting   $ \alpha=-1 $, we obtain
$
\delta \tilde{\mathcal{D}_0} =0. 
$
This means  that $ \delta W^{\mathrm{ren}}=0, $ and  there is no gravitational anomaly:
\beq
\langle \p_{\mu} T^{\mu}_{\,\,\, \nu} \rangle =0 \, .
\eeq
As a consequence, the divergence of the scale current in eq.~(\ref{Classical scale nonrel}) takes the form 
$
\langle \p_\mu J_S^\mu \rangle=
\langle T^i_{\,\,\, i} - 2 \epsilon^0\rangle  
$.
On the contrary, in Ref. \cite{Fernandes:2017nvx}, a non vanishing diffeomorphism anomaly was found using 
 Fujikawa's method, {\it i.e.}
\beq
\langle \p_{\mu} {(T_F)}^{\mu}_{\,\,\, \nu} \rangle = \frac12 \p_{\nu} \mathcal{A} \, ,
\eeq
 where ${\cal A }$ is the trace anomaly evaluated from the energy-momentum tensor ${(T_F)}^{\mu}_{\,\,\, \nu}$,
 regularized as in Ref. \cite{Fernandes:2017nvx}.
 This is not in contradiction with our result, as a ``subtracted'' energy momentum tensor, in the spirit of Ref.~\cite{Bardeen:1984pm},
  can be defined
 \beq
 \hat{T}^{\mu}_{\,\,\,\nu} = {(T_F)}^{\mu}_{\,\,\,\nu} - \frac12  \delta^{\mu}_{\,\,\,\nu} \mathcal{A} \, , 
\label{ridefinizione anomalia diffeo}
\eeq
in such a way it satisfies the conservation equation 
\beq
\langle \p_{\mu} \hat{T}^{\mu}_{\,\,\, \nu} \rangle  = 0 \, .
\eeq
It seems that the zeta function regularization method we used automatically selects the conserved energy momentum tensor ${T}^{\mu}_{\,\,\, \nu}=\hat{T}^{\mu}_{\,\,\, \nu}$. 
In fact, if we compare the trace of the subtracted $\hat{T}^{\mu}_{\,\,\, \nu}$ with eq.  (\ref{a4scalar}), we find  a substantial agreement  with \cite{Fernandes:2017nvx}\footnote{The disagreement concerns essentially  an overall sign. We were not able to figure out what is the origin of the discrepancy. }.


\section{Conclusions}
\label{conclu}

In this paper we found that a non-zero trace anomaly occurs
for a nonrelativistic scalar and fermion fields in $2+1$ dimensions,
 coupled to a particle number background $A_\mu$. This agrees with the results of \cite{Fernandes:2017nvx}.
 Analogous calculations \cite{Auzzi:2016lxb, Auzzi:2017jry}  performed in curved backgrounds without $A_\mu$ lead  to a result  proportional to the trace anomaly of a relativistic  scalar/fermion in 
$3+1$ dimensions. 
Instead, the anomaly in $A_\mu$ background 
is genuinely nonrelativistic, as it has no counterpart in the relativistic case.
The resulting  anomaly is not gauge-invariant in the background 
source vector field; similar non-gauge invariant anomalies are known to
 occur also in the relativistic case, see {\it e.g.} \cite{Bardeen:1969md} and
  \cite{Keren-Zur:2014sva}. 
  
We also computed the diffeomerphism anomaly for a scalar
and we found a vanishing result; this may be not in contradiction
with the results of  \cite{Fernandes:2017nvx}, because our energy-momentum
tensor may correspond to a subtracted version of the one studied in 
 \cite{Fernandes:2017nvx}.
 
Several open question are left for further investigation:
\begin{itemize} 
\item An analysis of the  Wess-Zumino consistency conditions
for trace anomalies in presence
of gauge and Milne boost violations would clarify the nature
 of the anomalies and their possible relevance for the properties of the RG flow.
Due to the large number of terms involved,
this seems a rather challenging task.  
 \item It would be interesting to study non relativistic
 anomalies for supersymmetric theories.
 In the  Schr\"odinger case it is likely that, in analogy to the relativistic case,
a non-trivial relation between superconformal R-charge  and trace anomaly exists
at the fixed point \cite{Anselmi:1997am,Intriligator:2003jj}.
 In the relativistic case, the traditional derivation 
 of these result relies on the equality
 of the flavor-$U(1)_R\,$-$U(1)_R$ and flavor-gravity-gravity triangle anomaly,
 due to supersymmetry.
  No axial anomaly is known in the 
 Schr\"odinger case\footnote{Axial anomaly instead was already
  studied in the Lifshitz case, see \cite{Bakas:2011nq}.}, and this makes
  the extension to the nonrelativistic case not straightforward.
  Another derivation of the relation between R-charges and $a-$anomaly
  was given in \cite{Auzzi:2015yia}, using the local RG approach by Osborn 
  \cite{Osborn:1991gm}; the relevant Wess-Zumino consistency conditions
  concerned terms in the trace anomaly which  are not 
  formally gauge-invariant \cite{Keren-Zur:2014sva}.
  It could be that a similar analysis might be extended to the nonrelativistic case. 
  \item An understanding of the case of anyons would be important
 for possible condensed matter applications. 
\end{itemize}


\section*{Acknowledgments}

We are grateful  to Arpita Mitra and Karan Fernandes for useful discussions.


\section*{Appendix}
\addtocontents{toc}{\protect\setcounter{tocdepth}{1}}
\appendix

\section{Weyl and Milne boost transformations}
\label{appeweyl}

The Newton-Cartan fields transform under Milne boosts in the following way:
\bea
v'^\mu & = & v^\mu+h^{\mu \nu} \psi_\nu \, \nl
h'_{\mu \nu} & = & h_{\mu \nu} -(n_\mu P_\nu^\rho+ n_\nu P_\mu^\rho) \psi_\rho
+n_\mu n_\nu h^{\rho \sigma} \psi_\rho \psi_\sigma \, , \nl
A'_\mu & = & A_\mu+P^\rho_\mu \psi_\rho -\frac{1}{2} n_\mu h^{\rho \s} \psi_\rho \psi_\s \, ,
\eea
where $ \psi_{\mu} $ is the local parameter of the trasformations. The fields $ n_{\mu} $ and $ h^{\mu\nu} $ are invariant.
These transformations are naturally implemented via null reduction technique.

The fundamental fields of Newton-Cartan geometry change as follow under 2+1 dimensional Weyl transformations:
\beq
n_{\mu} \rightarrow e^{2 \s} n_{\mu} \, , \qquad
v^{\mu}  \rightarrow e^{-2 \s} v^{\mu} \, , \qquad
h_{\mu\nu} \rightarrow e^{2 \s} h_{\mu\nu} \, , \qquad
h^{\mu\nu} \rightarrow e^{-2 \s} h^{\mu\nu} \, ,
\eeq
where $ \s $ is a local parameter and coordinates do not transform.
Note that the gauge field $ A_{\mu} $ is invariant.

The Newton-Cartan measure $ \sqrt{g} $ changes under 2+1 dimensional Weyl transormations as
\beq
\sqrt{\det(h_{\mu\nu}+n_{\mu}n_{\nu})}=\sqrt{g} \rightarrow e^{4 \s} \sqrt{g} \, .
\eeq

\section{Conventions on gamma matrices and vielbein}
\label{appegamma}

The $2+1$ dimensional dreibein $e^a_\mu $ is defined by dimensional reduction of 
the $3+1$ dimensional fierbein $e^A_M$:
\beq
e^A_{\,\,\, M} = \begin{pmatrix}
e^-_{\,\,\, M}  \\
e^+_{\,\,\, M}  \\
e^a_{\,\,\, M}
\end{pmatrix} = 
 \begin{pmatrix}
e^-_{\,\,\, -} \ \ & e^-_{\,\,\, \mu}  \\
e^+_{\,\,\, -} \ \ & e^+_{\,\,\, \mu} \\
e^a_{\,\,\, -} \ \ & e^a_{\,\,\, \mu}
\end{pmatrix}
=
 \begin{pmatrix}
1 \ \ & A_{\mu} \\
0 \ \ & n_{\mu} \\
\mathbf{0} \ \ & e^a_{\mu}
\end{pmatrix}
\, .
\eeq
The inverse vielbein  are:
\beq
e^M_{\,\,\, A} = \begin{pmatrix} e^M_{\,\,\, -}\ \ & e^M_{\,\,\, +} \ \ & e^M_{\,\,\, a}   \end{pmatrix} =
 \begin{pmatrix} e^-_{\,\,\, -}\ \ & e^-_{\,\,\, +} \ \ & e^-_{\,\,\, a}  \\
 e^{\mu}_{\,\,\, -}\ \ & e^{\mu}_{\,\,\, +} \ \ & e^{\mu}_{\,\,\, a}  \end{pmatrix}
= \begin{pmatrix}1 \ \ & -v^{\sigma} A_{\sigma} \ \ &  - h^{\nu \sigma} A_{\sigma} e^a_{\nu}  \\
\mathbf{0} \ \ &  v^{\mu} \ \ &   h^{\mu\nu} e^{a}_{\nu} \end{pmatrix} \, .
\eeq
The 2+1 dimensional dreibein $ e^{a}_{\,\,\, \mu} $ are not completely free, but they are related to other Newton-Cartan fields via the relations
\beq
e^M_{\,\,\, A} e^B_{\,\,\, M} = \delta_A^{\,\,\, B} \, , \qquad
e^A_{\,\,\, M} e^N_{\,\,\, A} = \delta_M^{\,\,\, N} \, .
\eeq

In order to deal with nonrelativistic fermions using the 
null-reduction method, it is necessary to introduce Pauli and Dirac matrices in 4 dimensions in light-cone indices.
The usual convention is
\beq
\sigma^{A} = (\mathbf{1}, \sigma^\a) \, , \qquad  \bar{\sigma}^{A} = (- \mathbf{1},  \sigma^\a) \, .
\eeq
In light-cone coordinates they become
\bea
\sigma^{\pm} = \frac{1}{\sqrt{2}} (\sigma^{3} \pm \sigma^{0}) \, , &&\qquad
\bar{\sigma}^{\pm} = \frac{1}{\sqrt{2}} (\bar{\sigma}^{3} \pm \bar{\sigma}^{0})  \, ,
\nl
\sigma^- = \sqrt{2}
\begin{pmatrix}
0 & 0 \\
0 & -1 
\end{pmatrix} \, , &&\qquad
\sigma^+ = \sqrt{2}
\begin{pmatrix}
1 & 0 \\
0 & 0 
\end{pmatrix} \, , 
\nl
\bar{\sigma}^- = \sqrt{2}
\begin{pmatrix}
1 & 0 \\
0 & 0 
\end{pmatrix} \, , && \qquad
\bar{\sigma}^+ = \sqrt{2}
\begin{pmatrix}
0 & 0 \\
0 & -1 
\end{pmatrix} \, ,
\nl
\sigma^1 = \bar{\sigma}^1=
\begin{pmatrix}
0 & 1 \\
1 & 0 
\end{pmatrix}  \, , && \qquad
\sigma^2 = \bar{\sigma}^2 =
\begin{pmatrix}
0 & -i \\
i & 0 
\end{pmatrix}  \, .
\eea
The associated Gamma matrices in 4 dimensions are
\bea
\gamma^{-}=\frac{1}{\sqrt{2}} (\gamma^{3}- \gamma^0) = \sqrt{2}  \begin{pmatrix}
0 & 0 &  0 & 0 \\
0 & 0 &  0 & -1 \\
1 & 0 &  0 & 0 \\
0 & 0 &  0 & 0 \\  
\end{pmatrix} \, , &&
 \,  \gamma^{+}= \frac{1}{\sqrt{2}} (\gamma^{3}+ \gamma^0) = \sqrt{2}  \begin{pmatrix}
0 & 0 &  1 & 0 \\
0 & 0 &  0 & 0 \\
0 & 0 &  0 & 0 \\
0 & -1 &  0 & 0 \\  
\end{pmatrix}  \, ,
\nl
\gamma^1 = \begin{pmatrix}
0 & \sigma^1 \\
 \sigma^1 & 0  
\end{pmatrix}  \, , && \qquad
\gamma^2 = \begin{pmatrix}
0 & \sigma^2 \\
 \sigma^2 & 0  
\end{pmatrix}  \, .
\eea
The corresponding Lorentz generators, needed to compute covariant derivatives, are:
\beq
\sigma^{AB} = \frac{1}{2} \left(  \sigma^A \bar{\sigma}^B  - \sigma^B \bar{\sigma}^A \right) \, , \qquad
\gamma^{AB}= \frac{1}{2} [\gamma^A , \gamma^B ] \, .
\eeq

\section{Time-independent insertion contributions to heat kernel}
\label{AppeA}

In this Appendix we will compute time-independent insertion of operators in the heat kernel expansion.
We derive the formula for a single insertion of the term $ a_i (x) $ multiplying a spatial derivative; all the other non-vanishing insertions can be found in \cite{Auzzi:2016lxb}-\cite{Auzzi:2017jry}.
We compute
\beq
K_{1 Q_i} (s) = \int_0^{s'} ds' \int d^d \tilde{x} \int d \tilde{t} \, \langle x t | e^{(s-s')\bigtriangleup}| \tilde{x} \tilde{t} \rangle 
Q_i (\tilde{x}) \frac{\p}{\p \tilde{x}_i}
\langle \tilde{x} \tilde{t} | e^{s' \bigtriangleup} | x' t' \rangle =
\eeq
\beq
= -  \frac{\p}{\p x'_i} \left[ \int_0^{s'} ds' \int d^d \tilde{x} \int d \tilde{t} \, \langle x t | e^{(s-s')\bigtriangleup}| \tilde{x} \tilde{t} \rangle Q_i (\tilde{x}) 
\langle \tilde{x} \tilde{t} | e^{s' \bigtriangleup} | x' t' \rangle   \right] \, ,
\label{single insertion a_i time independent}
\eeq
where we used  the parity properties of the flat solution of the heat kernel and of the derivative operation. In this way we recognize that the term in parenthesis is the definition of the single insertion of a term without derivatives of kind 
\beq
K_{1 P} (s)=  \int_0^{s'} ds' \int d^d \tilde{x} \int d \tilde{t} \, \langle x t | e^{(s-s')\bigtriangleup}| \tilde{x} \tilde{t} \rangle P (\tilde{x}) 
\langle \tilde{x} \tilde{t} | e^{s' \bigtriangleup} | x' t' \rangle \, ,
\eeq
where we only have to rename the operator as $ a_i (x) . $

The computation for this quantity was performed in \cite{Auzzi:2016lxb} and gave as a result
\beq
\begin{aligned}
K_{1 P} (s)  =  & \int_0^{s} ds'  \frac{1}{2 \pi^2} \frac{1}{(4 \pi s)^{d/2}} 
\frac{8 \pi m s }{4 m^2 s^2 + (t-t')^2}  \\
& \int \frac{d^d k}{(2 \pi)^{d/2}} \exp \le - \frac{(x-x')^2}{4s} + i k \cdot \le x \frac{s'}{s} + x' \frac{s-s'}{s}  \ri - k^2 \frac{s'}{s} (s-s')  \ri  a_i (k)  \, .
\end{aligned}
\eeq
If we now differentiate this expression with respect to $ x' $ and we put $ t=t', x=x' $ we obtain: 
\beq
\tilde{K}_{1 Q_i} (s) = \int_0^{s} ds'  \int \frac{d^d k}{(2 \pi)^{d/2}} \frac{2}{m (4 \pi s)^{d/2+1} } \le - i k_i \frac{s-s'}{s}  \ri \exp \le ikx - k^2 \frac{s'}{s} (s-s') \ri Q_i (k) \, .
\eeq
Performing the inverse Fourier transform and expanding around $ s=0 $ the exponential, we find the result
\beq
\tilde{K}_{1 Q_i} (s) = \frac{2}{m (4 \pi s)^{d/2+1} }\le - \frac{s}{2} \p_i Q_i (x) - \frac{s^2}{12} \p_i \p^2 Q_i (x) + \mathcal{O} (s^3) \ri \, .
\eeq

\section{Time-dependent insertion contributions to heat kernel}
\label{AppeB}
We want to generalize the results of the previous section concerning the heat kernel expansion by considering time-dependent insertions of operators.


\subsection{Single insertion computations}

Let us start with the single insertion of a term without derivatives acting on the fields. We consider the Fourier decomposition
\beq
P(x,t) = \int \frac{d^d k}{(2 \pi)^{d/2}} \int \frac{d \omega}{\sqrt{2 \pi}} \, P(k, \omega) e^{i(kx-\omega t)}  
\eeq
in order to find:
\beq
\begin{aligned}
K_{1 P} (s) = & \int_0^{s} ds' \int d^d \tilde{x} \int d \tilde{t} \, \langle x t | e^{(s-s')\bigtriangleup}| \tilde{x} \tilde{t} \rangle P (\tilde{x}, \tilde{t}) 
\langle \tilde{x} \tilde{t} | e^{s' \bigtriangleup} | x' t' \rangle = \\
=& \int_0^{s} ds'  \frac{1}{(2 \pi)^2}  \frac{1}{(4 \pi (s-s'))^{d/2}}   \frac{1}{(4 \pi s')^{d/2}}  \int \frac{d \omega}{\sqrt{2 \pi}} \int d \tilde{t} e^{-i \omega \tilde{t}} \frac{m(s-s')}{m^2 (s-s')^2 + \frac{(t-\tilde{t})^2}{4}} \\
 & \frac{m s'}{m^2 s'^2 + \frac{(\tilde{t}-t')^2}{4}}  \int \frac{d^d k}{(2 \pi)^{d/2}} e^{ik \tilde{x}} 
\exp \le - \frac{(x-\tilde{x})^2}{4(s-s')} - \frac{(\tilde{x}-x')^2}{4s'} \ri P(k, \omega)   \, ,
\end{aligned}
\label{general single insertion time dependent}
\eeq
where we used the explicit expression of the flat-space heat kernel for the Schr\"odinger operator, eq. (\ref{heatfree}).

We observe that the time and spatial parts of the integral decouple and appear as distinct multiplicative factors. We can then use the result for the spatial part that is presented in \cite{Auzzi:2016lxb}:
\beq
\begin{aligned}
& \int d^d \tilde{x} \int \frac{d^d k}{(2 \pi)^{d/2}} e^{ik \tilde{x}} 
\exp \le - \frac{(x-\tilde{x})^2}{4(s-s')} - \frac{(\tilde{x}-x')^2}{4s'}  \ri  = \\
& = \int \frac{d^d k}{2 \pi^{d/2}}  \exp \le - \frac{(x-x')^2}{4s} + i k \cdot \le x \frac{s'}{s} + x' \frac{s-s'}{s}  \ri - k^2 \frac{s'}{s} (s-s')  \ri   \, .
\end{aligned}
\eeq
In particular the $ x=x' $ result for the spatial part of the integral is
\beq
\int \frac{d^d k}{2 \pi^{d/2}}  \exp \le  i k \cdot x - k^2 \frac{s'}{s} (s-s')  \ri   \, .
\label{result of the spatial integration, time dep insertions}
\eeq
In order to compute the temporal part $ I(\omega) $ of the integral, we need to find the analytic structure in the complex plane of the integrand:
\beq
\begin{aligned}
I(\omega)= &  \int d \tilde{t} e^{-i \omega \tilde{t}} \frac{m(s-s')}{m^2 (s-s')^2 + \frac{(t-\tilde{t})^2}{4}} \frac{m s'}{m^2 s'^2 + \frac{(\tilde{t}-t')^2}{4}}  =  \\
= & 4 \alpha \beta  e^{-i \omega t'}  \int d \tilde{t} \frac{ e^{-i \omega \tilde{t}}}{(\tilde{t}+ i \beta)(\tilde{t}-i \beta)(\tilde{t} - \Delta t+ i \alpha)(\tilde{t} - \Delta t - i \alpha)} \, ,
\end{aligned}
\eeq
where we sent $ \tilde{t} \rightarrow \tilde{t} + t' $  and we defined
\beq
\alpha = 2m (s-s') \, , \qquad   \beta = 2ms'  \, , \qquad  \Delta t = t - t' \, .
\eeq
The quantities $ \alpha, \beta $ are positive by definition. We use the residue theorem to find
\beq
\begin{aligned}
I(\omega) & =   4 \alpha \beta  e^{-i \omega t'} \theta(\omega) \left[ \frac{\pi e^{- \beta \omega}}{\beta ((\Delta t + i \beta)^2 + \alpha^2)} +
 \frac{\pi e^{- \alpha \omega-i \Delta t \omega }}{\alpha ((\Delta t - i \alpha)^2 + \beta^2)}    \right] +  \\
& + 4 \alpha \beta  e^{-i \omega t'}  \theta (- \omega) \left[ \frac{\pi e^{\beta \omega}}{\beta ((\Delta t - i \beta)^2 + \alpha^2)} +
 \frac{\pi e^{ \alpha \omega-i \Delta t \omega }}{\alpha ((\Delta t + i \alpha)^2 + \beta^2)}    \right]  \, .
\end{aligned}
\eeq
It can be found that the expression for $ \omega=0 $ gives the time-independent results found in \cite{Auzzi:2016lxb}-\cite{Auzzi:2017jry} if we choose the prescription $ \theta(0)= 1/2 $ for the Heaviside distribution:
\beq
I(0) =\frac{8\pi m s}{4 m^2 s^2 + (t-t')^2}=  \int d \tilde{t} \frac{m(s-s')}{m^2 (s-s')^2 + \frac{(t-\tilde{t})^2}{4}} 
  \frac{m s'}{m^2 s'^2 + \frac{(\tilde{t}-t')^2}{4}} \, .
\eeq
The heat kernel computation only requires equal-time insertions, then we put $ \Delta t =0 $ to obtain
\beq
I(\omega, \Delta t=0) =  \frac{2 \pi}{m s} \frac{1}{s-2s'} \left[ e^{-2m s' |\omega|} (s-s') - s' e^{-2m(s-s') |\omega|}  \right] =
\frac{2 \pi}{m s} + \mathcal{O} (s) \, .
\label{equal time insertion time dependent integral}
\eeq
Using eqs. (\ref{result of the spatial integration, time dep insertions}) and (\ref{equal time insertion time dependent integral}) inside 
eq. (\ref{general single insertion time dependent}) and expanding in the auxiliary time \emph{s} we finally obtain the result
\beq
\tilde{K}_{1 P} (s)= \frac{2}{m (4 \pi s)^{d/2+1}} \le s P(x,t) + \frac{1}{6} s^2 \p_i^2 P(x,t) + \mathcal{O} (s^3)   \ri \, .
\eeq
This is the same result of the case without time-dependence because the first order of the expansion of exponential terms vanishes.

Let us now consider the single insertion of an operator with a spatial derivative acting on the fields. Once again, we find that eq. (\ref{single insertion a_i time independent}) is satisfied and then the calculation reduces to applying a spatial derivative to the single insertion $ K_{1 P} (s) : $
\beq
K_{1 Q_i } (s)= -  \frac{\p}{\p x'_i} \left[ \int_0^{s'} ds' \int d^d \tilde{x} \int d \tilde{t} \, \langle x t | e^{(s-s')\bigtriangleup}| \tilde{x} \tilde{t} \rangle Q_i (\tilde{x}, \tilde{t}) 
\langle \tilde{x} \tilde{t} | e^{s' \bigtriangleup} | x' t' \rangle   \right] \, .
\eeq
Since the expression in parenthesis does not change if we add a time dependence to the operators of the heat kernel expansion, and since spatial and temporal parts of the integral factorize, we obtain an equivalent formula also for
\beq
\tilde{K}_{1 Q_i} (s) = \frac{2}{m (4 \pi s)^{d/2+1} }\le - \frac{s}{2} \p_i Q_i (x,t) - \frac{s^2}{12} \p_i \p^2 Q_i (x,t) + \mathcal{O} (s^3) \ri \, .
\eeq
Single insertions of operators with a time derivative applied to the dynamical fields $ S(x,t) $ can be modified by time dependence, but since they vanish on our background we will not consider this kind of terms.


\subsection{Double insertion computations}
We consider the double insertion of operators of kind $ P(x,t) , $ which is given by 
{\small \beq
\begin{aligned}
 & K_{2 P} (s)  =  \int_0^s ds_2 \int_0^{s_2} ds_{1} \int d^d x_1 \int d^d x_2 \int dt_1  \int dt_2  \, \langle x t |  e^{(s-s_2)\bigtriangleup} | x_2 t_2 \rangle  & \\
 & \qquad \qquad P(x_2, t_2) \langle x_2 t_2 |  e^{(s_2-s_1)\bigtriangleup} | x_1 t_1 \rangle  P(x_1, t_1) \langle x_1 t_1 | e^{s_1 \bigtriangleup} | x' t' \rangle =  \\
 & =  \int_0^s ds_2 \int_0^{s_2} ds_{1} \frac{1}{(2\pi)^3} \frac{1}{(4 \pi (s-s_2))^{d/2}} \frac{1}{(4 \pi (s_2-s_1))^{d/2}} \frac{1}{(4 \pi s_1)^{d/2}} \int d^d x_1 \int d^d x_2 \int dt_1  \int dt_2 & \\
&   \int \frac{d^d k_1}{(2 \pi)^{d/2}}  \int \frac{d^d k_2}{(2 \pi)^{d/2}} 
\exp \le ik_1 x_1 + i k_2 x_2 - \frac{(x'- x_2)^2}{4(s-s_2)} - \frac{(x_2-x_1)^2}{4(s_2 - s_1)} - \frac{(x_1-x)^2}{4 s_1}  \ri  \int \frac{d \omega_1}{\sqrt{2 \pi}}\int \frac{d \omega_2}{\sqrt{2 \pi}} & \\
&    e^{-i \omega_1 t_1 - i \omega_2 t_2} \frac{m(s-s_2)}{m^2 (s-s_2)^2 + \frac{(t-t_2)^2}{4}} \frac{m (s_2-s_1)}{m^2 (s_2-s_1)^2 + \frac{(t_2- t_1)^2}{4}}  \frac{m s_1}{m^2 s_1^2 + \frac{(t_1-t')^2}{4}} P(k_2, \omega_2) P(k_1, \omega_1) & \, .
\end{aligned}
\label{espansione inserzione doppia di P}
\eeq }
It is evident that also in this situation the time and spatial parts of the integral factorize. The latter was found in \cite{Auzzi:2016lxb}  to be
{\small \[
\Xi (x,x') = \int d^d x_1 \int d^d x_2 
\exp \le  -\frac{(x'-x_2)^2}{4 (s-s_2)}
 -\frac{(x_2-x_1)^2}{4 (s_2-s_1)} 
-\frac{(x_1-x)^2}{4 s_1}+ i k_1 x_1+i k_2 x_2 \ri
\]
\[
= (4 \pi)^{d} \le \frac{s_1 (s-s_2) (s_2-s_1)}{s} \ri^{d/2}
\exp \le 
\frac{i k_1 s_1 x'}{s}+\frac{i k_2 s_2 x'}{s}-\frac{i k_1 s_1 x}{s}
-\frac{i k_2 s_2 x}{s}+\frac{k_1^2 s_1^2}{s}+\frac{k_2^2 s_2^2}{s}
\right.
\]
\[ 
\left.
-k_1^2   s_1-2 k_1 k_2 s_1-k_2^2 s_2+\frac{2 k_1 k_2 s_1 s_2}{s}+i k_1 x+i k_2 x
-\frac{x^2}{4 s}+\frac{x x'}{2 s}-\frac{\left(x'\right)^2}{4 s} \ri  \, .
\] }
At coincident points it becomes
\[
\Xi (x,x)
= (4 \pi)^{d} \le \frac{s_1 (s-s_2) (s_2-s_1)}{s} \ri^{d/2} 
\]
\beq
\exp \le  ik_1 x_1 + i k_2 x_2 +
k_1^2 \le \frac{s_1^2}{s} - s_1 \ri + k_2^2 \le \frac{s_2^2}{s} - s_2 \ri +2 k_1 k_2 \le \frac{s_1 s_2}{s} - s_1 \ri  \ri \, .
\label{parte spaziale inserzione doppia}
\eeq
Now we analyze the temporal part:
 \bea
\Psi (t,t', \omega_1, \omega_2)& &=  \int dt_1 \int dt_2  e^{-i \omega_1 t_1 - i \omega_2 t_2} \nonumber\\
&&\!\!\! \!\!\!   \frac{m(s-s_2)}{m^2 (s-s_2)^2 + \frac{(t-t_2)^2}{4}} \frac{m (s_2-s_1)}{m^2 (s_2-s_1)^2 + \frac{(t_2-t_1)^2}{4}}  \frac{m s_1}{m^2 s_1^2 + \frac{(t_1-t')^2}{4}} \, .
\eea 
The integral in the variable $ t_1 $ can be performed using the previous technique. If we set
\beq
\alpha = 2m (s_2 - s_1) \, , \qquad   \beta = 2m s_1 \, , \qquad  \Delta t = t_2 - t' \, ,
\eeq
we find
\beq
\begin{aligned}
  I(\omega_1) & = \int dt_1   e^{- i \omega_1 t_1}\frac{m (s_2-s_1)}{m^2 (s_2-s_1)^2 + \frac{(t-\tilde{t})^2}{4}}  \frac{m s_1}{m^2 s_1^2 + \frac{(t-\tilde{t})^2}{4}} = \\
 & =   4 \alpha \beta  e^{-i \omega t'} \theta(\omega_1) \left[ \frac{\pi e^{- \beta \omega_1}}{\beta ((\Delta t + i \beta)^2 + \alpha^2)} +
 \frac{\pi e^{- \alpha \omega-i \Delta t \omega }}{\alpha ((\Delta t - i \alpha)^2 + \beta^2)}    \right] +  \\
& + 4 \alpha \beta  e^{-i \omega t'}  \theta (- \omega_1) \left[ \frac{\pi e^{\beta \omega}}{\beta ((\Delta t - i \beta)^2 + \alpha^2)} + \frac{\pi e^{ \alpha \omega-i \Delta t \omega }}{\alpha ((\Delta t + i \alpha)^2 + \beta^2)}    \right]  \, .
\end{aligned}
\eeq
The last step in the time integration consists in evaluating
\beq
\Psi (t,t', \omega_1, \omega_2) = \int dt_2 e^{-i \omega_2 t_2} \frac{m(s-s_2)}{m^2 (s-s_2)^2 + \frac{(t-t_2)^2}{4}} I (\omega_1) \, .
\eeq
The result of the evaluation is very cumbersome, but  it can be checked that, assuming the prescription $\theta(0)=1/2$, it  gives the exact time-independent result in the limit of vanishing frequencies:
\beq
\Psi (t=t', \omega_1=\omega_2=0) = \frac{16 \pi^2 \theta^2(0)}{m s} = \frac{4 \pi^2}{m s}  \, .
\eeq
Moreover, in order to compute the insertions of time-dependent operators we only need the lowest orders of the expansion in \emph{s} of the solution at coincident points, which turns out to be 
\beq
\Psi (t=t', \omega_1, \omega_2) =  \frac{4 \pi^2}{m s} e^{-i (\omega_1 + \omega_2)t} + \mathcal{O}  (s) \, .
\label{parte temporale inserzione doppia}
\eeq
The zeroth order in the variable \emph{s} vanishes.

Combining eqs. (\ref{parte spaziale inserzione doppia}) and (\ref{parte temporale inserzione doppia}) into (\ref{espansione inserzione doppia di P}) we find the same result of the time-independent case:
\beq
\tilde{K}_{2PP} = \frac{2}{m(4 \pi s)^{d/2+1}} \left( \frac{s^2}{2} P(x,t)^2 + \mathcal{O} (s^3) \right)  \, .
\eeq
Additional new terms will contribute only to higher orders in \emph{s}, and therefore they do not modify the $ a_4 $ coefficient.

Since time and space integrals factorize and there are no contributions to lower-order terms in the heat kernel expansion, we can similarly find that $ \tilde{K}_{2 X} $ have the same expressions of the time-independent case, if we choose among the set
\beq
X= \lbrace P(x,t) , Q_i (x,t)  \rbrace \, .
\eeq
Additional terms could appear in insertions concerning the operator $ S(x,t)  $. They  will not be considered here because $ S(x,t) $ vanishes in all the backgrounds studied in this paper.

\end{document}